\documentclass[conference]{IEEEtran}
\IEEEoverridecommandlockouts
\usepackage{cite}
\usepackage{amsmath,amssymb,amsfonts}
\usepackage{algorithmic}
\usepackage{graphicx}
\usepackage{textcomp}
\usepackage{xcolor}
\usepackage{multirow} 
 \usepackage{booktabs}
 \usepackage{graphicx} 
 \usepackage{graphicx, caption, sidecap}
\usepackage{caption}
\usepackage{amssymb}
\def\BibTeX{{\rm B\kern-.05em{\sc i\kern-.025em b}\kern-.08em
    T\kern-.1667em\lower.7ex\hbox{E}\kern-.125emX}}
\usepackage{array, makecell} %
\usepackage{stmaryrd}

\begin{document}

\title{Learning on JPEG-LDPC Compressed Images: Classifying with Syndromes}


\author{
Ahcen Aliouat and Elsa Dupraz \\
\small  IMT Atlantique, CNRS UMR 6285, Lab-STICC, Brest, France \thanks{ This work has received a French government support granted to the Cominlabs excellence laboratory and managed by the National Research Agency in the `Investing for the Future' program under reference ANR-10-LABX-07-01.}
}
\maketitle

\begin{abstract}
In goal-oriented communications, the objective of the receiver is often to apply a Deep-Learning model, rather than reconstructing the original data. 
In this context, direct learning over compressed data, without any prior decoding, holds promise for enhancing the time-efficient execution of inference models at the receiver. However, conventional entropic-coding methods like Huffman and Arithmetic break data structure, rendering them unsuitable for learning without decoding. In this paper, we propose an alternative approach in which entropic coding is realized with Low-Density Parity Check (LDPC) codes. We hypothesize that Deep Learning models can more effectively exploit the internal code structure of LDPC codes. At the receiver,  we leverage a specific class of Recurrent Neural Networks (RNNs), specifically Gated Recurrent Unit (GRU), trained for image classification.  Our numerical results indicate that classification based on LDPC-coded bit-planes surpasses Huffman and Arithmetic coding, while necessitating a significantly smaller learning model. This demonstrates the efficiency of classification directly from LDPC-coded data, eliminating the need for any form of decompression, even partial, prior to applying the learning model.
\end{abstract}


\section{Introduction}
Over the past few years, learning on coded data has emerged as an important area of research due to the increasing amount of images and videos that need to be processed in current and next-generation networks. 
This emerging paradigm is often referred to as Image and Video Coding for Machines (VCM) \cite{duan2020video}.
A key desired feature of this paradigm is that the receiver may apply the considered learning task without any prior decoding of the data. Such a feature would present a compelling advantage for the time-efficient execution of fundamental learning tasks such as classification, segmentation, or recommendation, directly on compressed data. 

In this context, this work considers the important case of classification over compressed images.
Recently, full end-to-end Deep Learning approaches have been proposed for this purpose, where the receiver applies learning on a well-designed latent space \cite{torfason2018towards,duan2023unified,zhang2023machine}. However, these approaches often lack the capability to reconstruct images when necessary, and they deviate from compliance with established standards for image compression, such as JPEG.
Alternatively, in~\cite{hill2021transform,pistono2020training,9956532,fu2016using}, learning methods are applied over JPEG-compressed images. Yet, these methods necessitate partial decoding steps, such as reconstructing the Discrete Cosine Transform (DCT) coefficients, before the application of the learning task.
In addition, some other studies, such as~\cite{chamain2021end},  aim to apply machine vision or object segmentation over compressed videos. These approaches encounter similar limitations as the methods proposed for images, either lacking compatibility with prevalent video coding standards like MPEG or HEVC or mandating partial decoding before task execution.

Interestingly,~\cite{remy1,piau2023predicting} make a step toward overcoming these limitations by investigating classification directly over entropy-coded images. These works encompass usual entropy coding techniques, such as Huffman and Arithmetic coding. They demonstrate that established Deep Learning models for image classification, such as ResNet or VGG, can be directly applied to entropy-coded images without necessitating any partial decoding.  However, this comes at a cost of reduced classification accuracy, attributed to the disruptive nature of conventional entropy coders on image structure, notably compromising spatial closeness between pixels. 

In this paper, our objective is to explore alternative entropy-coding techniques that offer greater relevance from a learning standpoint while retaining the capability for data reconstruction. Specifically, we concentrate on Low-Density Parity Check (LDPC) codes, a family of channel codes that have also demonstrated efficiency for entropic coding at large~\cite{caire2004noiseless} as well as in distributed source coding~\cite{liveris2002compression}, and the context of 360-degree images~\cite{9319718}. Our rationale for examining this unconventional entropic-coding approach stems from the key idea that its inherent structure may be leveraged more efficiently by Deep Learning models for image classification.

\begin{figure*}[htbp]
\includegraphics[width=\linewidth]{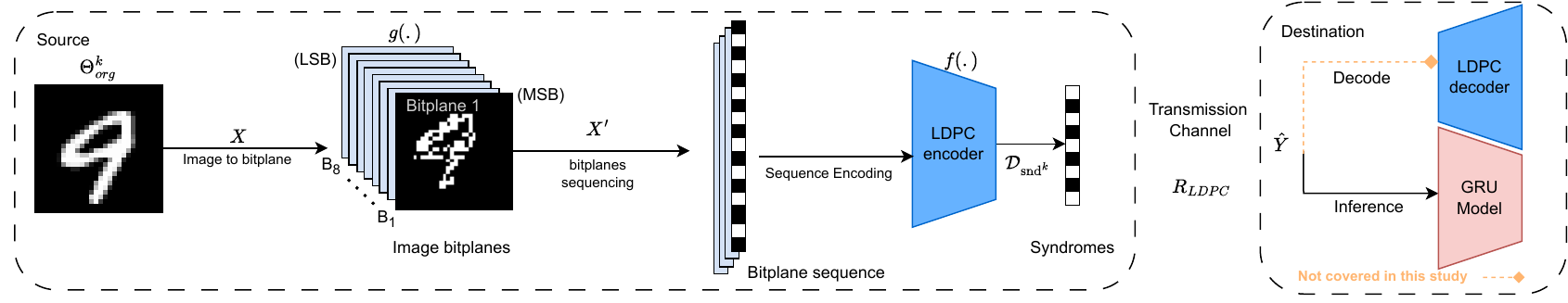}
\caption{\textbf{First setup}: Syndromes obtained from the LDPC-coded bit-planes are fed as input of a GRU model for classification}
\label{setup1}
\end{figure*}

We consider standard JPEG compression, with its typical steps such as DCT and quantization, while substituting the final entropic coder with LDPC coding applied over bit planes. The later step generates distinct codewords, termed syndromes, for each bit-plane. At the receiver, we employ Recurrent Neural Networks (RNN), specifically the Gated Recurrent Unit (GRU) model, to classify images based on their LDPC-coded bit-plane syndromes. This choice of RNN draws inspiration from the work of~\cite{bennatan2018deep}, which initially demonstrated the ability of RNN structures in approaching Maximum-Likelihood channel decoding of linear block codes.

Our experimental results demonstrate 
the feasibility of applying image classification directly over LDPC-coded syndromes without the need for any partial decoding. Furthermore, this approach yields superior accuracy compared to classification over Huffman or Arithmetic entropic coding, while employing models of significantly reduced complexity. 
This approach has the potential to streamline the learning pipeline by eliminating the need for decoding. Additionally, it offers a new perspective on the synergy between compression techniques and Deep Learning models.

The remainder of the paper is organized as follows. Section~\ref{sec:jpeg-ldpc} describes JPEG-compressed images with entropic-coder based on LDPC codes. Section~\ref{sec:classif} introduces the proposed classification method. Section~\ref{SCM} details the experiments conducted. Section~\ref{sec:results} presents and discusses the results.

\section{JPEG Coding with LDPC Codes}\label{sec:jpeg-ldpc}
This section briefly describes JPEG compression and shows how LDPC codes can be used for bit-plane entropy-coding in this context.

\subsection{JPEG Coding}
The conventional JPEG compression workflow starts with a DCT transform, followed by a quantization step. Subsequently, Zig-Zag scanning as well as Run-Length encoding are applied to rearrange the DCT coefficients, enhancing data suitability for entropic coding. Finally, Huffman or Arithmetic entropy-coding methods are employed to minimize the size of the compressed data based on symbol probability. For a comprehensive understanding of JPEG compression, readers are referred to \cite{125072}.

The works presented in~\cite{remy1,piau2023predicting} show that image classification over Huffman or Arithmetic coded data is possible. However, this comes at the price of a significant loss in classification accuracy, due to the fact that such entropic coding methods break the semantic and closeness structure of the pixels. Nevertheless, these properties are essential for usual Deep Learning models dedicated to image classification. This motivates the need to explore other entropic-coding methods that are more appropriate for learning on coded data.

\subsection{JPEG Coding with LDPC Codes}
In this work, we propose to use LDPC codes for entropic coding. Indeed, we hypothesize that the LDPC coder may be more relevant for image classification, since it may better preserve the structure of the images features, through the LDPC code structure. Further, it was shown in~\cite{caire2004noiseless}  that LDPC codes are appropriate for entropic-coding, at the price of a slightly higher coding rate. 
To effectively implement this approach, we now describe the specific requirements for applying binary LDPC codes to $n$ non-binary pixels or quantized DCT coefficients. First, we transform the successive non-binary symbols into $I$ bit-planes, where the bit-plane with the most significant bits (MSBs) contains the majority of an image's information.
Previous studies have demonstrated that LDPC bit-plane encoding is practical and does not significantly impact the coding rate compared to non-binary LDPC encoding~\cite{9319718}. 

 \begin{figure*}[htbp]
\includegraphics[width=\linewidth]{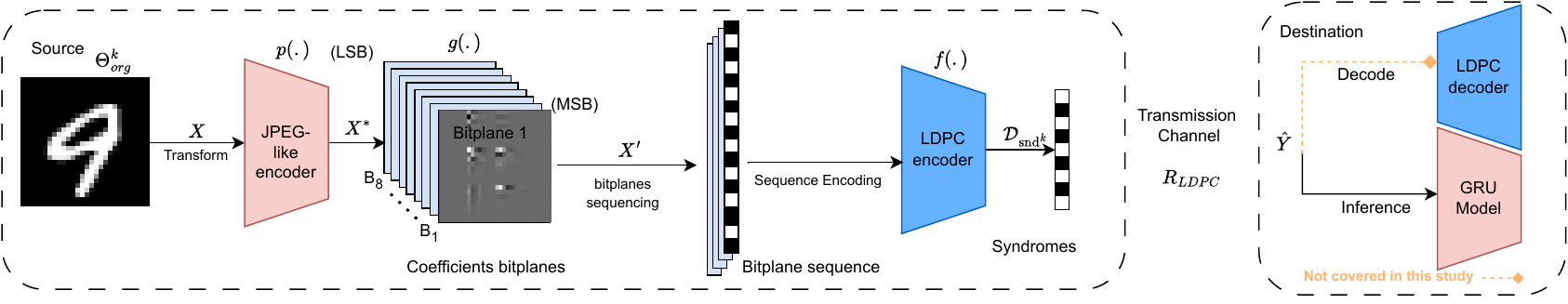}
\caption{\textbf{Second setup}: Syndromes of the DCT-LDPC coefficients bit-planes are fed as inputs of a GRU model for classification}
\label{setup2}
\end{figure*}

Next, let $H$ be the binary parity check matrix of size $m \times n$, where
($m < n$) of a binary LDPC code~\cite{liveris2002compression,ye2019optimized}. If $H$ is full rank,
the source coding rate is defined as $R = m/n$. The parity check matrix $H$ can be equivalently described as a Tanner graph, which is bipartite in nature, connecting $n$ source nodes with $m$ syndrome nodes. 
In our scheme, each bit plane $\mathbf{x}_i$, $i \in \llbracket 1,I\rrbracket$ has the same length $n$ and it is compressed by an LDPC code using the following formula~\cite{liveris2002compression}:
\begin{equation}\label{eq:ldpc_enc}
    \mathbf{s}_i = H\mathbf{x}_i
\end{equation}
where $\mathbf{s}_i$ is a binary sequence of length $m$ called the syndrome. In our scheme, the $I$ syndromes will be sent to the decoder as the compressed information. At the receiver, these syndromes will be used either for data reconstruction as in~\cite{9319718}, or as inputs to a learning model.

\section{Classification over LDPC-coded Images}\label{sec:classif}

In this section, we describe the GRU model we consider for classification over LDPC-coded images. 

\subsection{GRU for compressed image sequences classification}
The GRU model simplifies the Long Short Term Memory (LSTM) framework by combining the forget and input gates into a single update gate and introducing a reset gate, thereby reducing complexity without significantly sacrificing performance. Unlike LSTMs, GRUs manage the flow of information without using a separate memory cell, simplifying the processing of temporal sequences.
Consider a GRU with \(J\) units. The activation variable \(h_{t}^{j}\) for the unit \(j\) of the GRU at time \(t\) is computed as a weighted average between the previous activation variable \(h_{t-1}^{j}\) and the candidate activation variable \(\tilde{h}_{t}^{j}\), as follows:

\begin{equation}
h_{t}^{j} = \left(1 - z_{t}^{j}\right) h_{t-1}^{j} + z_{t}^{j} \tilde{h}_{t}.
\end{equation}
This equation simplifies the information update process within the unit.
The update gate \(z_{t}^{j}\) determines the extent to which the unit updates its activation and is calculated using:
\begin{equation}
z_{t}^{j} = \sigma\left(W_{z} \boldsymbol{u}_{t} + U_{z} \boldsymbol{h}_{t-1}\right),
\end{equation}
where $\mathbf{u}_t$ is the input signal, and \(\sigma(\cdot)\) is the sigmoid activation function, 
Furthermore, the candidate activation \(\tilde{h}_{t}^{j}\) is computed like the update gate, but incorporates the reset gate \(r_{t}^{j}\) to potentially discard information that is no longer relevant:
\begin{equation}
\tilde{h}_{t}^{j} = \tanh\left(W \boldsymbol{u}_{t} + U \left(r_{t}^{j} \odot \boldsymbol{h}_{t-1}\right)\right),
\end{equation}
where \(\odot\) denotes an element-wise multiplication. Here, \(r_{t}^{j}\) represents a set of reset gates, and \(W_z\), \(W\), \(U_z\), and \(U\) are the weights associated with the update gate and the candidate activation respectively.

\begin{figure}[htbp]
\centering \includegraphics[width=0.6\linewidth]{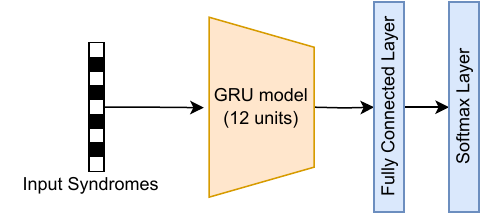}
\caption{Considered learning model for classification }
\label{model}
\end{figure}

The GRU model introduced by Cho et al. \cite{cho2014properties}, includes a reset gate \(r_{t}^{j}\) that allows for the selective forgetting of prior information. This capability simulates the scenario of encountering the initial element of an input sequence whenever the reset gate is inactive (\(r_{t}^{j}=0\)). Such a mechanism enables the model to dynamically adapt its memory focus. The reset gate's operation is governed by the equation:
\begin{equation}
r_{t}^{j} = \sigma\left(W_{r} \boldsymbol{u}_{t} + U_{r} \boldsymbol{h}_{t-1}\right).
\end{equation}

The update gate \(z_t^j\) plays a pivotal role in modulating the influence of previous states on the current state, allowing the model to balance between short-term and long-term dependencies. This balance is achieved through the combined activity of the reset and update gates. The model involving the considered GRU architecture is summarized in Fig. \ref{model}.

\subsection{Learning setups}\label{sec:learning_setups}

We consider an initial dataset of images $\Theta_{\text{org}}$, and investigate two setups. 
In the first setup, we do not consider JPEG compression: for each image $\boldsymbol{\theta}_k \in \Theta_{\text{org}}$, the pixels are directly transformed into bit-planes and encoded using LDPC codes. The $I$ resulting syndromes $\mathbf{s}_{k,i}$, $i \in \llbracket 1,I\rrbracket $, form a new data $\mathbf{d}_k^{(1)}$. We refer to the corresponding new dataset as $\mathcal{D}_{\text{snd}}^{(1)}$, where the generation of one element $\mathbf{d}_k^{(1)}$ of the dataset $\mathcal{D}_{\text{snd}}^{(1)}$ can be described as follows:
\begin{equation}
\mathbf{d}_k^{(1)} = f(g(\boldsymbol{\theta}_k)) .    
\end{equation}

Here, the function $g(.)$ represents the bit-plane constructor, and the function $f(.)$ is the LDPC source encoding according to~\eqref{eq:ldpc_enc}.

For the second setup, we mimic JPEG compression and first apply the DCT transform followed by standard JPEG quantization. We do not consider Zig-Zag scanning nor RLE (since they were specific to Huffman coding) and directly transform the DCT coefficients into bit-planes. In this case, each element $\mathbf{d}_k^{(2)}$ of the new dataset  $\mathcal{D}_{\text{snd}}^{(2)}$ is obtained as
\begin{equation}
    \mathbf{d}_k^{(2)} = f(g(p(\boldsymbol{\theta}_k))) .  
\end{equation}
The function $p(.)$ represents the JPEG-like operation, which carries out $8\times8$ DCT transform and quantization. For the quantization step, we employ the standard JPEG quantization matrix. The quality factor is fixed as $\text{QF} = 95$.

\begin{table}[htbp]
    \caption{Hyper-parameters for the learning model}

    \centering
    \begin{tabular}{p{0.2\columnwidth}|p{0.15\columnwidth}|p{0.25\columnwidth}|p{0.2\columnwidth}}
     \toprule
        \textbf{parameter} & \textbf{Value} & \textbf{parameter} & \textbf{Value} \\ \midrule
        Learning rate &  0.001 & Dropout rate & 0.2 \\  
        No. epochs & 30 & L2 Regularization & 0.002  \\  
        No. GRU Units & 12 (32) & Activation function & Softmax \\ 
        Batch size & 64 &Optimizer & Adam  \\ 
         \bottomrule
        
    \end{tabular}
    \label{tab:learning_param}
\end{table}

\begin{table*}[htbp]
\centering
\caption{Classification accuracy of the GRU on LDPC coded data compared to the state-of-the-art for multiple datasets.}
\label{tab:performance_metrics}
\begin{tabular}{llcccccccccc}
\toprule
\textbf{Dataset} & \textbf{Model} & \multicolumn{2}{c}{\textbf{No coding}} & \multicolumn{3}{c}{\textbf{Coding on Orig. (Setup1)}} & \multicolumn{5}{c}{\textbf{Coding on JPEG (Setup2)}} \\ 
\cmidrule(lr){1-1} \cmidrule(lr){2-2} \cmidrule(lr){3-4} \cmidrule(lr){5-7} \cmidrule(lr){8-12} 
 &  & \textbf{None} & \textbf{\makecell{None\\MSB}} & \textbf{Huff\cite{remy1}} & \textbf{Arith\cite{remy1}} & \textbf{LDPC} & \textbf{JPEG\cite{remy1}} & \textbf{\makecell{DCT\\-tr.\cite{fu2016using}}} & \textbf{\makecell{J-L\\8bp}} & \textbf{\makecell{J-L\\MSB}} &\textbf{\makecell{J-L\\MSB+1bp}}\\ 
\midrule
\multirow{2}{*}{\textbf{\textit{MNIST}}} &\textbf{GRU12(proposed)} & 0.9439 & 0.8842 & - & - & 0.8192  & - & - & 0.9060 & 0.6548&0.8791\\
                       &\textbf{GRU32(proposed)} & 0.9799 & 0.9154 & - & - & 0.8556  & - & - & 0.9237 & 0.6843&0.8849\\
                              \cmidrule(lr){2-12}
                       & \textbf{VGG11 \cite{remy1}} & 0.9891 & - & 0.8323 & 0.6313 & - & -  & - & - & -&-\\
                       & \textbf{URESNET18 \cite{remy1}} & 0.9875 & - & 0.7450 & 0.5949 & - & -  & - & - & -&-\\
                              \cmidrule(lr){2-12}
                       & \textbf{FullyConn \cite{fu2016using}} & 0.9200 & - & - & - & -  & - & $0.9000$ & - & -&-\\
\midrule
\multirow{2}{*}{ \makecell{\textbf{\textit{Fashion}} \\ \textbf{\textit{-MNIST}}}} &\textbf{GRU12} & 0.8616 & 0.8052 & - & - & 0.8166  & - & - & 0.8332 & 0.5222&0.8325\\
                               &\textbf{GRU32} & 0.8750 & 0.8314 & - & - & 0.8306  & - & - & 0.8434 & 0.5395&0.8414\\
                                      \cmidrule(lr){2-12}
                               & \textbf{VGG11 \cite{remy1}} & 0.9018 & - & 0.7634 & 0.6898 & - & -  & - & - & -&-\\
                               & \textbf{URESNET18 \cite{remy1}} & 0.8497 & - & 0.6862 & 0.6116 & - & -  & - & - & -&-\\
\midrule
\multirow{2}{*}{\makecell{\textbf{\textit{YCIFAR}}\\\textbf{\textit{-10}}}} &\textbf{GRU12} & 0.3127 & 0.3249 & - & - & 0.4023  & - & - & 0.4234 & 0.1350&0.3537\\
                           &\textbf{GRU32} & 0.3596 & 0.3560 & - & - & 0.4023  & - & - & 0.4316 & 0.1403&0.3544\\
                                  \cmidrule(lr){2-12}
                           & \textbf{VGG11 \cite{remy1}} & 0.5657 & - & 0.3606 & 0.2976 & -  & 0.3245 & - & - & -&-\\
                           & \textbf{URESNET18 \cite{remy1}} & 0.3836 & - & 0.2591 & 0.2432 & -  & - & - & - & -&-\\
                                  \cmidrule(lr){2-12}
                           & \textbf{FullyConn \cite{fu2016using}} & 0.3800 & - & - & - & -  & - &  0.3000$ $ & - & -&-\\
\bottomrule
\multicolumn{12}{l}{\scriptsize *J-L: JPEG-LDPC, MSB+1bp: Sign bit-plane of the DCT coefficients + the next bit-plane after the sign bit-plane} 
\end{tabular}

\end{table*}

The two datasets $\mathcal{D}_{\text{snd}}^{(1)}$ and $\mathcal{D}_{\text{snd}}^{(2)}$ will serve as inputs to two GRU learning models for training. In what follows, we aim to quantitatively evaluate the efficiency of applying such classification models directly onto LDPC-coded syndromes, without any prior decoding.

\section{Data and Experiments}\label{SCM}
We now describe the considered datasets, as well as the experimental setups for our performance evaluation. The experiments are divided into two parts, aligned with the considered two setups. The first set of experiments evaluates the ability of the GRU model to learn from the LDPC syndromes of one or multiple bit-planes of the image. The second set of experiments investigates the model classification performance over the considered JPEG-like chain where Huffman coding is replaced by LDPC coding. 

\subsection{Datasets}
The tests are conducted on three standard datasets: MNIST \cite{lecun1998gradient}, Fashion-MNIST \cite{xiao2017fashion}, and CIFAR-10 \cite{krizhevsky2009learning}.
For MNIST and Fashion-MNIST, images are grayscale-converted and processed as outlined in Section~\ref{sec:learning_setups}. While CIFAR images are converted to YCrCb, only the Y channel is considered for coding, benefiting from its ability to retain most information from the original RGB images.

\subsection{Code parameters}

In our experiments, we consider a regular $(3,6)$-LDPC parity-check matrix $H$ of size $1024\times512$ with rate $1/2$. 
In order to perform a consistent evaluation, the same parity-check matrix is used in all the experiments. Accordingly, the sizes of the images of the three datasets are resized from $28 \times 28$ to $32 \times 32$ by padding zeros at the end of the columns and rows. This adjustment also facilitates the application of the $8\times8$ DCT transform. 

\subsection{Model parameters}

In the experiments, the GRU model's input size is set to $512 \times I$ when applied over the $I$ LDPC-coded syndromes, and to $1024 \times I$ when applied over the original $I$ bit-planes. The latter corresponds to learning from the uncoded data, and it is considered for comparison purposes. For GRU model training, datasets are divided into training/validation sets at ratios of $84\%$-$16\%$ for MNIST and Fashion-MNIST, and $86\%$-$14\%$ for YCIFAR-10. The used hyper-parameters are shown in Table~\ref{tab:learning_param}.

\subsection{Rate gain}
We propose to measure the achieved gain while considering LDPC bit-planes coding using the following equation:
\begin{equation}
     \Gamma = \frac{R_{\text{LDPC}}}{N_{\text{bp}}}.
\end{equation}
Where $R_{\text{LDPC}}$ is the LDPC coding rate, and $N_{bp}$ is calculating as the number $I$ of considered bit-planes divided by the maximum number of bit-planes $\eta_{bp}$, ie: $N_{bp}=I/\eta_{bp}$.  

\section{Results and Discussion}\label{sec:results}
In this section, the GRU model's performance is reported in terms of classification accuracy and model complexity.

\subsection{Accuracy comparison}
Table~\ref{tab:performance_metrics} shows the classification accuracy over LDPC-coded syndrome under different conditions, including: (i) without coding for reference, (ii) by applying LDPC coding directly on the original image bit-planes (Setup $1$), (iii) with JPEG-like compression (Setup $2$). Results are compared to those of Huffman and Arithmetic coding methods from ~\cite{remy1}, and to the "truncated DCT" technique of~\cite{fu2016using}. 
For Setup $1$, learning is applied over the $I=8$ bit-planes. For Setup $2$, we also report on the impact of the number of bit-planes on the model performances, when learning over $1$ bit-plane (J-L MSB), when considering $2$ bit-planes (J-L MSB + $1$ bp), and when considering the $8$ bit planes (DL). It is worth noting that the results of~\cite{piau2023predicting}
 for Huffman and Arithmetic coding also considered learning over the $8$ bit-planes.
It is remarkable that \textit{Setup $1$} significantly outperforms other methods in accurately classifying entropic-coded images. We observe a performance improvement of about $15\%$ for the CIFAR-10, $10\%$ for Fashion-MNIST, and MNIST, compared to the best results of the previously mentioned studies.
The results also reveal that \textit{Setup 2}, which combines DCT and quantization with LDPC coding, surpasses the performance of \textit{Setup 1}, which solely relies on LDPC coding, as shown in Table \ref{tab:performance_metrics}. This superiority most probably stems from the DCT's efficacy in feature representation through signal frequencies. Consequently, integrating DCT with LDPC coding enhances syndrome classification accuracy, highlighting the proposed GRU model's performance improvements due to the synergistic effect of these techniques.
Furthermore, examining the impact of utilizing limited bit-planes (Columns J-L-MSB and J-L-MSB+bp) reveals their unexpected efficiency. While learning from only the MSB bit-plane $(I=1)$ degrades the performance, learning from the MSB bit-plane and one additional bit-plane $(I=2)$ yields results nearly equivalent to using all original bit-planes. This finding shows the feasibility of compressing data through selective bit-plane usage without compromising learning efficacy.
\subsection{Accuracy-Complexity balance}

According to the complexity analysis shown in Table~\ref{tab:acc_weights}, our proposed method requires only $19k$ learnable parameters to outperform 2D-CNN, 1D-CNN, and "$1024$ Fully Connected" models. When using GRU on compressed images as feature sequences instead of image pixels (or coefficients), complexity reduction is about $1600\%$, which is significant.
This confirms the inherent ability of LDPC codes to preserve the structure of the data, which significantly simplifies the learning process for the model, compared to Huffman and Arithmetic coding.

\begin{table}[htbp]
    \centering
    
    \begin{tabular}{lcc}
    \toprule

       \textbf{Setup} &  \textbf{Best accuracy} & \textbf{No. of learnables}\\
        \midrule
        Huffman on VGG \cite{remy1}& 0.8323 & 138M\\
        Huffman on Resnet \cite{remy1}& 0.6313 & 60M \\
      \cmidrule(lr){1-3}
        DCT on FullyConnected\cite{fu2016using} & 0.9000 & 1M \\
     \cmidrule(lr){1-3}
        JPEG-like LDPC on 12 units GRU& \textbf{0.9060}  & \textbf{19k}\\
        JPEG-like LDPC on 32 units RGU& \textbf{0.9237}  & \textbf{52.6k}\\
        \bottomrule
        
    \end{tabular}
    \caption{Reported Accuracy vs weights for MNIST dataset}
    \label{tab:acc_weights}
\end{table}

\subsection{Compression gain}
The same LDPC parity-check matrix, with a rate $R_{\text{LDPC}}=1/2$ was used in all the experiments. By encoding only one bit-plane (MSB) ($N_{\text{bp}}=1/8$), the compression ratio can reach up to $0.5$ bits per pixel. On the other hand, encoding a maximum of $8$ bit-planes yields a compression ratio of $4$ bits per pixel.
It is noteworthy that incorporating a pruning step before LDPC coding can significantly improve the compression gain. However, its impact on accuracy also warrants investigation.

\section{Conclusion}
In this paper, we have proposed a lightweight learning model based on GRU for learning from LDPC source entropy-coded data. We have demonstrated the efficiency of this approach in terms of classification accuracy and model complexity. As a result, integrating LDPC codes within a JPEG-like chain better preserves the data structure, and opens the way to learning over coded data, without any prior decoding. Future work will include evaluating the impact on the learning performance of JPEG quality factors, the number of bit planes used for learning, and pruning strategies, which all change the compression ratio. We will also investigate the performance of various regular and irregular LDPC codes in such learning contexts and explore the idea of designing LDPC codes specifically for learning from syndromes. 

\bibliographystyle{ieeetr}
\bibliography{biblio}

\end{document}